\documentclass{sigchi-ext}
\usepackage[T1]{fontenc}
\usepackage{textcomp}
\usepackage[scaled=.92]{helvet} % for proper fonts
\usepackage{graphicx} % for EPS use the graphics package instead
\usepackage{balance}  % for useful for balancing the last columns
\usepackage{booktabs} % for pretty table rules
\usepackage{ccicons}  % for Creative Commons citation icons
\usepackage{ragged2e} % for tighter hyphenation

\def\plaintitle{Tensions on Trails: Understanding Differences between Group and Community Needs in Outdoor Settings} 
\def\emptyauthor{}
\def\plainkeywords{Hiking, Trails, Wilderness, Groups, Communities}

\title{Tensions on Trails: Understanding Differences between Group and Community Needs in Outdoor Settings}

\numberofauthors{4}
\author{%
  \alignauthor{%
    \textbf{Lindah Kotut}\\
    \affaddr{Department of Computer Science and Center for HCI}\\    
    \affaddr{Virginia Tech} \\
    \affaddr{Blacksburg, VA 24061, USA} \\
    \email{lkotut@vt.edu} }\alignauthor{%
    \textbf{Shuo Niu}\\
    \affaddr{Department of Computer Science and Center for HCI}\\   
    \affaddr{Virginia Tech} \\
    \affaddr{Blacksburg, VA 24061, USA} \\
    \email{shuoniu@vt.edu} } \vfil \alignauthor{%
    \textbf{Michael Horning}\\
    \affaddr{Department of Communication and Center for HCI}\\    
    \affaddr{Virginia Tech} \\
    \affaddr{Blacksburg, VA 24061, USA} \\
    \email{mhorning@vt.edu} } \alignauthor{%
    \textbf{Timothy Stelter}\\
    \affaddr{Department of Computer Science and Center for HCI}\\ 
    \affaddr{Virginia Tech} \\
    \affaddr{Blacksburg, VA 24061, USA} \\
    \email{tstelter@vt.edu} } \vfil \alignauthor{%
    \textbf{Derek Haqq}\\
    \affaddr{Department of Computer Science and Center for HCI}\\ 
    \affaddr{Virginia Tech} \\
    \affaddr{Blacksburg, VA 24061, USA} \\
    \email{dhaqq@vt.edu} } \alignauthor{%
    \textbf{D. Scott McCrickard}\\
    \affaddr{Department of Computer Science and Center for HCI}\\   
    \affaddr{Virginia Tech} \\
    \affaddr{Blacksburg, VA 24061, USA} \\
    \email{mccricks@cs.vt.edu} } 
    }

\definecolor{linkColor}{RGB}{6,125,233}
\hypersetup{%
  pdftitle={\plaintitle},
  pdfauthor={\emptyauthor},
  pdfkeywords={\plainkeywords},
  bookmarksnumbered,
  pdfstartview={FitH},
  colorlinks,
  citecolor=black,
  filecolor=black,
  linkcolor=black,
  urlcolor=linkColor,
  breaklinks=true,
}

\begin{document}

%% For the camera ready, use the commands provided by the ACM in the Permission Release Form.
\CopyrightYear{2018}
\setcopyright{rightsretained}
%\conferenceinfo{Workshop on HCI Outdoors: Understanding Human-Computer Interaction in the Outdoors at CHI 2018}{'18 Montréal, Canada} 
%\isbn{0-12345-67-8/90/01}
\doi{http://dx.doi.org/10.1145/2858036.2858119}
%% Then override the default copyright message with the \acmcopyright command.
%\copyrightinfo{\acmcopyright}

		\maketitle

% Uncomment to disable hyphenation (not recommended)
% https://twitter.com/anjirokhan/status/546046683331973120
\RaggedRight{} 

% Do not change the page size or page settings.

%============================= ABSTRACT =============================
\begin{abstract}
This paper compares the needs of groups and communities in outdoor settings, seeking to identify subtle but important differences in the ways that their needs can be supported.   We first examine the questions of who uses technology in outdoor settings, what their technological uses and needs are, and what conflicts exist between different trail users regarding technology use and experience. We then consider selected categories of people to understand their distinct needs when acting as groups and as communities. We conclude that it is important to explore the tensions between groups and communities to identify design opportunities. 
\end{abstract}



%============================= INTRODUCTIONS ========================
\section{Introduction}
Rural areas lack some of the capabilities and amenities that are found in urban areas, but they often have vast wilderness spaces for hiking and other outdoor activities that long have been touted as an enriching, and worthy of preservation and even cultivation \cite{Nash14}.  However, not everyone has the same objectives when using wilderness spaces, leading to differing goals even among people who are identified as part of the same collection of people. There are assumptions that are necessarily made about the outdoor spaces: who uses trails, what technology they use, and their attitudes toward the usage of said technologies \cite{Bryson1998}. Tension can exist in the roles of groups in outdoor settings. Hunters, for example, agree on the ethos of ``fair chase" \cite{Su2017}, but different types of hunters differ on how they interpret this notion depending on their attitude towards the role of weapon technology (crossbows vs bows, rifles vs muzzleloaders, rifles vs bows) in hunting. The role of technology enhances personal experience on the trail, such as the use of fitbits and headphones \cite{Anderso2017}; citizen scientist water quality monitoring \cite{Rapousis2015}, and logistical planning of trail practicalities (e.g., campsite reservations, rest-room facilities).

%--------------------- TABLE --------------------
\begin{table*}
  \centering
  \begin{tabular}{l r r r r r}
    \toprule
Activists &Families &Historians &Mental/Physical Rehab &Property Owners &Section Hikers\\
Bikers &Farmers &Hunters &Park Rangers &Recreational &Sponsored Hikers\\
Bird Watchers &Firemen &Horse-Back Riders &Pet Owners &Retirees &Thru Hikers\\
Boy/Girl Scouts &Fishermen &Locals  &Picnickers &Solo Hikers &Tourists\\
Day Hikers &Guide-Book Authors &Loggers &Plant Foragers &Scientists &Trail Angels\\
Exercisers &Herbalists  &Maintenance Workers &Prof/ Army Training &Search \& Rescue & \\
    \bottomrule
  \end{tabular}
  \caption{35 unique collections of people were curated from previously identified hiker roles and used to study technological opportunities. }~\label{tab:table1}
\end{table*}
%-------------------- END TABLE ----------------

This paper looks at the differences in the concerns of two different collections of people: groups and communities.  \textit{Groups} tend to be small, focused, and somewhat exclusive, with membership centered on some sort of agreed-upon criteria such as voting, leader confirmation, or birth, that may exclude certain people. Members of a group tend to have some familiarity with each other, toward establishing common (or at least compatible) goals and approaches to situations. In comparison, \textit{communities} generally are larger and center on common beliefs, concerns, or behaviors. Communities share some common characteristics, but they may have different beliefs and approaches to addressing their commonalities.  A half dozen colleagues that share an office would be considered a group, while people might live together in a gated \textit{community}.

Differences between these terms have been long debated.  Grudin's classic CSCW paper \cite{grudin1994computer} does not explicitly define these terms, but it does refer to groups as a subset of an organization that tend to be small and task-focused, while referring to communities as larger and loosely connected around ideas and themes (e.g., the CSCW community, the ESS community, the R\&D community).  Daniel Ospina differentiates groups as having a sense of belonging and shared purpose, while communities may share the belonging but may differ in their practices and values \cite{Ospina17}. We can also look to tech industries for the distinction between these terms. A Facebook group is an invitation-based collection of people that share specific interests or backgrounds, while a Facebook community is open to anyone with an expressed interest in a topic. Once you are in a Facebook group, you have a great deal of power to post, comment, and invite others, while a Facebook community has a leadership structure that control information flow.

%============================= APPROACH ============================

Based on this minimal pair of definitions for group and community, this paper sought to examine how select trail-related collections of people differed in their goals and approaches.  This approach builds upon previous affinity diagramming sessions \cite{Kotut2017,Kotut2018} used to identify different facets of roles and goals for technology on the trail. The first session involved 25 participants tasked with identifying \textit{who} the trail users are, while the second affinity diagramming session involving 9 participants was used to determine \textit{why} these people are on the trail based on common goals identified through clustering, from which 35 emerged (Table \ref{tab:table1}). As the main contribution of this position paper, six HCI faculty and grad students discussed these collections and identified some of particular interest within the group-community perspective.

%============================= OBSERVATIONS ==========================
\section{Observations}
As part of a group session, we examined the differences between group and community concerns for each of the collections of people in Table 1.  Our initial goal was to categorize them primarily into one of the two, but instead we found great interest in the tensions between groups and communities for each of the collections of people (e.g., activists exist as both groups and communities).  This section defines several categories of hikers and details key discussion points.

{\bf Thru hikers}\\
Hiking as an activity can be done for recreation, wellness and fitness, competition, experiencing nature, and more. Hikers in the United States also tend to avoid urban areas and seek to embrace the wilderness \cite{Bryson1998}. The \textit{thru hiker} has the goal of completing a chosen trail in its entirety within one hiking season. An attempt of the 2,190 mile Appalachian Trail (AT) for example, would take several months to complete in one hiking season. 
% Given this extreme nature of thru-hiking, personal goals such as self reflection, competitive nature, getting away from society etc., are often cited to be reasons governing the choice of attempting a thru hike. 

Community tends to be important to \textit{thru hikers}, given the numerous Facebook pages, blogs, planned meet-ups, and other social media and activities that are customarily used before, during and after the hike.  Additionally, many hikers embrace the notion of a \textit{group} with fellow hikers -- often seeking to camp together, share meal planning, and splitting the weight of tents and cookware.  In addition to this, \textit{thru hikers} are also known to remain connected with other groups (e.g., families, co-workers) and find ways to maintain those social bonds even while undertaking the hike.

Despite the commonality of the overarching ``thru hike'' goal, conflict arises between groups on issues of preferences such as taste in (or absence of) music, communication styles etc.

{\bf Exercisers}\\
Examples of trail users considered as \textit{exercisers} include: \textit{Day-Walkers, Bikers, Joggers} and \textit{Horse-back riders}. Although this classification emerged through consideration of shared goals associated with exercises and training: losing weight, muscle building, endurance training, general fitness, or simply as a means of deriving personal fulfillment and pleasure, a commonality may be derived from the environment/place that such individuals choose to use; i.e. the choice of outdoors as a means of engaging in exercise activity over indoor alternatives. This suggests that \textit{exercisers} seek to fulfill specific wants, needs, and goals that may be either interconnected or independent of the activity and exercise-related goals. Perceptions of enhanced enjoyment, fulfillment, motivation, sense of peace, solitude and/or richer stimulus, may all be reasons why such individuals opt to use the outdoors, because users have interrelated, competing, and sometimes conflicting, priorities in terms or want and need fulfillment.    

As such, a number of perspectives may emerge. For a given exercise, one may argue that these individuals may be viewed to be a \textit{group} or a \textit{community} depending on their level of involvement, interaction and commitment.  A great many devices and exercise programs leverage group behaviors, either cooperatively or competitively. 
%[MORE TO DISCUSS HERE]
From the community perspective, some shared norms, behavior, or culture related to preserving individual community member's ``sense'' of the outdoor medium emerge and influence the actions of the members.  This may include an increased awareness of, and respect for, the outdoor exercise experience of a fellow member of the community of outdoor exercisers.  A member of such a community may be more aware of the effect of intrusive stimuli; e.g., noise pollution from a jogger listening to music without the aid of headphones. % or headset, on the experience of another
As such, they are likely to engage in their activity in a manner that preserves the sense of place.

%%% TBD?
% Understanding a user's competing priorities, perhaps through some formal multi-criteria decision analysis, would...something about design lol 

{\bf Activists}\\
\textit{Activists} as trail-users emerged from the original exercise when considering people who care about the trail, in combination with short-term goals (e.g., proper trash disposal) and long-term strategies (e.g., sustainability, such as activists protesting pipeline construction impacting the trail\cite{Appalachian18}). In reviewing our prior clustering activity, participants considered the trail-user placement on a social scale based on willingness to socialize, where \textit{anti-social} and \textit{extremely social} emerged as opposing extremes on the axis. Trail users tending toward the \textit{extremely social} end of the scale, were labeled as ``people to meet while hiking on the trail" \cite{Kotut2017}, and the \textit{activists'} nature of promoting/protesting actions on the trail and the inevitability of encounters with other trail users lend them towards this \textit{group} categorization. 

However, goals of activists can differ. Does the need of those who call for the preservation of the integrity of the trail usurp those of who lobby for economic benefits from a pipeline? This possibility of sharply opposing goals seems to preclude the categorization of \textit{activists} as a \textit{community}. Diverging goals aside however, what we find to be common concerns across different activist groups are issues of \textit{reach} -- given what definition emerged during the activity and considering examples form the trail \cite{Appalachian18}, \textit{mobilization}.

%============================= FUTURE PLANS =========================
\section{Conclusions and Future Plans}
Before designing for the trail, we argue that we need to understand the users and the tensions they experience. This position paper posits that differences within fairly well defined professions, hobbies, and activities can highlight conflicts.  We expect that workshop feedback will help guide our directions toward crafting a richer set of categories. 

\section{About the Authors}
\textbf{Lindah Kotut} is a Ph.D. student in the Department of Computer Science at Virginia Tech with interests in mobile devices and the risk inherent in using them. \\ 
\textbf{Michael Horning} is an Assistant Professor in the Department of Communication at Virginia Tech whose research focuses on social and psychological effects of communication technologies.\\  
\textbf{Derek Haqq} is a Ph.D. student in the Department of Computer Science at Virginia Tech and a IT management consultant and educator.\\
\textbf{Shuo Niu} is a Ph.D. student at Virginia Tech focusing on using large display technologies to understand social media like tweets and blogs.\\  
\textbf{Tim Stelter }is a Ph.D. student in the Department of Computer Science focusing on understanding outdoor communities on  trail adventures.\\  
\textbf{Scott McCrickard }is an Associate Professor in the Department of Computer Science at Virginia Tech and the founder of the Tech on the Trail initiative, which examines the impacts of technology on extended outdoor situations.

\balance{} 

\bibliographystyle{SIGCHI-Reference-Format}
\bibliography{extended-abstract}

\end{document}